\documentclass [preprint,aps,amsmath,amssymb]{revtex4-1}

\usepackage{graphicx}
\usepackage{dcolumn}
\usepackage{bm}
\usepackage{color}
\usepackage{ulem}

\begin{document}


\title{Dirac dynamical resonance states around Schwarzschild black holes}


\author{Xiang-Nan Zhou}
\email{zhouxn10@lzu.edu.cn}
\affiliation{Institute of Theoretical Physics, Lanzhou University, Lanzhou 730000, People's Republic of China}

\author{Xiao-Long Du}
\email{duxl11@lzu.edu.cn}
\affiliation{Institute of Theoretical Physics, Lanzhou University, Lanzhou 730000, People's Republic of China}

\author{Ke Yang}
\email{yangk09@lzu.edu.cn}
\affiliation{Institute of Theoretical Physics, Lanzhou University, Lanzhou 730000, People's Republic of China}

\author{Yu-Xiao Liu \footnote{Corresponding author}}
\email{liuyx@lzu.edu.cn}
\affiliation{Institute of Theoretical Physics, Lanzhou University, Lanzhou 730000, People's Republic of China}



\begin{abstract}
Recently, a novel kind of scalar wigs around Schwarzschild black holes---scalar dynamical resonance states were introduced in [Phys. Rev. D 84, 083008 (2011)] and [Phys. Rev. Lett. 109, 081102 (2012)]. In this paper, we investigate the existence and evolution of Dirac dynamical resonance states. First we look for stationary resonance states of a Dirac field around a Schwarzchild black hole by using the Schr\"{o}dinger-like equations reduced from the Dirac equation in Schwarzschild spacetime. Then Dirac pseudo-stationary configurations are constructed from the stationary resonance states. We use these configurations as initial data and investigate their numerical evolutions and energy decay. These dynamical solutions are the so-called ``Dirac dynamical resonance states". It is found that the energy of the Dirac dynamical resonance states shows an exponential decay. The decay rate of energy is affected by the resonant frequency, the mass of Dirac field, the total angular momentum, and the spin-orbit interaction. In particular, for ultra-light Dirac field, the corresponding particles can stay around a Schwarzschild black hole for a very long time, even for cosmological time-scales.
\end{abstract}

\maketitle  


\section{Introduction}
\label{section1}

The ``no-hair" theorem postulates that all stationary black holes in general relativity can be completely characterized by only three externally observable classical parameters: the mass of the black hole $M$, { angular momentum $J$, and electric charge $Q$} \cite{Misner:1973}. But it is difficult to find isolated black holes in the real world, many people have faith in that we can get more details about black holes. Even though ``no-hair" conjecture forbids any invariable, stationary field configurations around the black holes, it says nothing about the lifetime of these fields' evolutions. One of the most famous investigations of the properties of various fields around black holes is the study of ``quasinormal modes", which is derived by considering appropriate boundary conditions \cite{PhysRevD.22.2323,Ohashi:2004wr}. It was argued that one can find the direct evidence of the existence of a black hole by comparing the observational data of gravitational wave with the quasinormal modes {from theoretical} arithmetic. Furthermore, the loop quantum gravity also supports the existence of quasinormal modes \cite{PhysRevLett.81.4293,PhysRevLett.90.081301}. But {quasinormal modes are} not the only kind of field configurations that can stay around black holes for a long time.

{Recently, a new method was proposed to get long-lasting scalar field configurations which are related to the effective potential well in the equation of motion \cite{PhysRevD.84.083008,PhysRevLett.109.081102}.} This kind of configurations are called ``dynamical resonance states" \cite{PhysRevLett.109.081102} or ``quasi-bound states" \cite{Degollado:2013eqa}, and they have complex frequencies with the real part representing the oscillation and the imaginary part representing the rate of decay.
{The ``dynamical resonance states" are associated with ``quasinormal modes" but satisfy different boundary conditions.
In Ref. \cite{PhysRevLett.109.081102}, the authors discussed the differences among the stationary resonances, the quasiresonances
(which come from the quasinormal modes), and the dynamical resonance states. It was shown that both the stationary resonances and quasiresonances are in fact nonphysical solutions because their energy densities both diverge at the horizon. But in the case of dynamical resonance states, the energy densities do not diverge. Furthermore, the frequencies of oscillation of the dynamical resonance states coincide with that of the stationary resonances, while the decay rates coincide with the imaginary part of the frequencies of quasiresonant modes. To make it more perspicuous, their differences and relations are shown in table \ref{tab3}.}
This new kind of solutions have been investigated in different types of black hole spacetimes \cite{PhysRevD.86.104026,Babichev:2012sg,Magana:2012xe,Degollado:2013eqa}.

On the other hand, most investigations of fields around a black hole {mainly} focus on the scalar field because of its extensive applications in modern cosmology \cite{PhysRevD.33.1590,PhysRevD.55.7546,PhysRevD.62.024027,PhysRevD.63.124015,PhysRevD.64.084017,PhysRevLett.88.151301,PhysRevD.66.044009,PhysRevD.66.084007,PhysRevD.66.124013,PhysRevD.68.044024,PhysRevD.67.024003,Setare:2003hm,PhysRevD.69.044016,PhysRevD.86.083531,PhysRevLett.109.111101,Dolan:2012yt,PhysRevD.85.024036,PhysRevD.87.043513}. In contrast, the study of the Dirac field around a black hole is relatively few. {The early investigations of Dirac quasinormal modes were done with the WKB analysis} \cite{PhysRevD.68.024003,Cho:2004wj}. Then the P\"{o}shl-Teller potential approximation was introduced \cite{PhysRevD.30.295,PhysRevD.71.124006,Jing:2005ux}. Although the Dirac quasinormal modes in different black hole spacetimes and the late-time evolution of the charged massive Dirac fields have been considered \cite{PhysRevD.72.027501,Lasenby:2002mc,Gibbons:2008a,Moderski:2008nq,Gibbons:2008gg,Chakrabarti:2008xz,Dolan:2009kj,R:2008rq,Wang:2009hr,LopezOrtega:2010uu,Cubrovic:2010bf,Varghese:2010qv,Oikonomou:2012js}, there is not any discussion about Dirac dynamical resonance states at present. Noting that the Dirac field has more complex structure and more freedoms than the scalar field, it motives us to consider these questions: Do there exist Dirac dynamical resonance states in the Schwarzschild black hole spacetime? If they do exist, how long will they stay around the Schwarzschild black hole? Will the spin and charge affect the Dirac dynamical resonance states and their {lifetimes} of evolution? Therefore, we pay our attention to the Dirac dynamical resonance states in the Schwarzschild black hole spacetime and discuss these questions in this paper.

Arrangement for this paper is as follows. In Sec.~\ref{section2}, we briefly review the Dirac equation in Schwarzschild spacetime and reduce it to a set of Schr\"{o}dinger-like equations. Then, in Sec.~\ref{section3}, we discuss the stationary resonance states of Dirac field using these Schr\"{o}dinger-like equations. In Sec.~\ref{section4}, we construct pseudo-stationary configurations from the stationary solutions and use them as initial data to perform the numerical evolutions. Finally, the conclusions and discussions are given in Sec.~\ref{section5}.

\begin{widetext}
\begin{center}
\begin{table}[htbp]
\caption{{Differences and relations among stationary resonances, quasiresonances and dynamical resonances. Here $\psi$ is the radial part of the matter field, $\omega$ is the frequency, and $m$ is the mass of the matter field. $r_{h}$ is the horizon radius and $r^{*}$ is the radial coordinate in the ``tortoise" coordinate system. Note that for quasiresonances, the frequency $\omega$ is complex.\label{tab3}}}
\begin{tabular}{lccc}
\hline\hline
${}$ & $~~~~~r^{*}\rightarrow-\infty(r\rightarrow r_{h})~~~~~$ & $~~~~~r^{*}\rightarrow+\infty(r\rightarrow+\infty)~~~~~$ & ~~~~~Energy~~~~~ \\
\hline
Stationary resonances     & $\psi\sim \text{e}^{-\text{i}k r^{*}}+\text{e}^{\text{i}k r^{*}}$ & $\psi\sim\text{e}^{-kr^{*}}$ & Divergent\\
${}$                      & $(k=\sqrt{m^2-\omega^2})$ & ${}$ & (Nonphysical)\\
\hline
Quasiresonances           & $\psi\sim \text{e}^{-\text{i}\sigma r^{*}}$  & $\psi\sim \text{e}^{\text{i}\sigma r^{*}}$        & Divergent\\
${}$                        & $(\sigma=\sqrt{\omega^2-m^2})$ & ${}$ & (Nonphysical)\\
\hline
Dynamical resonances      & $\psi(r<r_h+\epsilon)=0$ & $\psi\sim\text{e}^{-kr^{*}}$                                    & Finite\\
${}$                      &~~~~~~($\epsilon$ is a small positive constant)~~~& ${}$ & (Physical)\\
\hline\hline
\end{tabular}
\end{table}
\end{center}
\end{widetext}

\section{Dirac equation in Schwarzschild spacetime}
\label{section2}

We consider a test spinor field around a Schwarzschild black hole, which means that back raction will not be considered. The Dirac equation in a general spacetime is given by
\begin{eqnarray}
\left(\Gamma^{\mu}\mathfrak{D}_{\mu}+m\right)\Psi=0, \label{Dirac equation 1}
\end{eqnarray}
where $m$ is the mass of the Dirac field and $\mathfrak{D}_{\mu}=\partial_{\mu}+\Omega_{\mu}$ represents the covariant derivative. The spin connection $\Omega_{\mu}$ is defined by
{
\begin{eqnarray}
\Omega_{\mu}&=&\frac{1}{8}[\gamma^a,\gamma^b]e_{a}^{~\nu}e_{b\nu;\mu}, \\
e_{b\nu;\mu}&=&\partial_{\mu}e_{b\nu}-\Gamma^{\alpha}_{\mu\nu}e_{b\alpha}.
\end{eqnarray}}
$e_{a}^{~\mu}$ is the vierbein and satisfies $\eta_{ab}=g_{\mu\nu}e_{a}^{~\mu}e_{b}^{~\nu}$. In this whole paper, we shall use the positive signature, $\eta_{ab}=\text{diag}(-,+,+,+)$. $\gamma^{a}$ are gamma matrices in flat spacetime and satisfy $\left\{\gamma^{a},\gamma^{b}\right\}=2\eta^{ab}$. Specifically, we choose the Dirac representation:
\begin{eqnarray}
\gamma^0=\left( \begin{array}{cc}
     -\text{i}\mathbb{I}   & \mathbb{O} \\
      \mathbb{O}   & \text{i}\mathbb{I}
\end{array} \right),\qquad
\gamma^i=\left( \begin{array}{cc}
    \mathbb{O}    & -\text{i}\sigma^i \\
    \text{i}\sigma^i      & \mathbb{O}
\end{array} \right).
\end{eqnarray}
Then $\Gamma^{\mu}=\gamma^{a}e_{a}^{~\mu}$ are the gamma matrices in a general spacetime and satisfy $\left\{\Gamma^{\mu},\Gamma^{\nu}\right\}=2g^{\mu\nu}$.

For the Schwarzschild spacetime, the metric can be written as
\begin{eqnarray}
ds^2=-N(r)dt^2+\frac{dr^2}{N(r)}+r^2\left(d\theta^2+\sin^2\theta d\phi^2\right).
\end{eqnarray}
Here $N(r)=1-\frac{2M}{r}$ with $M$ the mass of the black hole. This coordinate system is suitable in the exterior region $r\in(2M,\infty)$. In addition, we choose the following vierbein for convenience \cite{PhysRevD.68.024003,PhysRevD.66.124017}:
{\begin{eqnarray}
e_{a}^{~\mu} = \left(
\begin{array}{cccc}
\frac{1}{\sqrt{N}}&0&0&0\\
0 & \sqrt{N}\sin\theta\cos\phi & \frac{\cos\theta\cos\phi}{r} & -\frac{\csc\theta\sin\phi}{r} \\
0 & \sqrt{N}\sin\theta\sin\phi & \frac{\cos\theta\sin\phi}{r} &  \frac{\csc\theta\cos\phi}{r} \\
0 & \sqrt{N}\cos\theta & -\frac{\sin\theta}{r} & 0 \\
\end{array}\right).~~~
\end{eqnarray}}
Then, the Dirac equation (\ref{Dirac equation 1}) can be written as \cite{PhysRevD.66.124017,Cho:2004wj,Brill:1957ab}:
{\begin{eqnarray}
  \frac{\gamma^{0}\partial_{t}\Psi}{\sqrt{N}}
    +\frac{\tilde{\gamma}}{r}N^{\frac{1}{4}}\partial_{r}(rN^{\frac{1}{4}}\Psi) -\frac{\tilde{\gamma}}{r}(\vec{\Sigma}\cdot\vec{L}+1)\Psi+m\Psi=0, ~~~ \label{Dirac equation 2}
\end{eqnarray}}
where $\vec{\Sigma}$ is defined as
\begin{eqnarray}
\vec{\Sigma}=\left( \begin{array}{cc}
     \vec{\sigma}    & \mathbb{O} \\
    \mathbb{O}       & \vec{\sigma}
\end{array} \right),
\end{eqnarray}
$\vec{L}$ is the standard angular momentum operator, and $\tilde{\gamma}=\gamma^{1}\sin\theta\cos\phi+\gamma^{2}\sin\theta\sin\phi+\gamma^{3}\cos\theta$. Defining
\begin{eqnarray}
\Psi(t,r,\theta,\phi)=\frac{1}{rN^{\frac{1}{4}}}\text{e}^{-\text{i}\omega t}\Phi(r,\theta,\phi) \label{Psi}
\end{eqnarray}
with $\omega$ a real frequency, Eq. (\ref{Dirac equation 2}) can be simplified as
\begin{eqnarray}
  -\frac{\text{i}\omega\gamma^{0}}{\sqrt{N}}\Phi
  +\tilde{\gamma}\sqrt{N}\partial_{r}\Phi
  -\frac{\tilde{\gamma}}{r}\left(\vec{\Sigma}\cdot\vec{L}+1\right)\Phi+m\Phi=0. ~~~ \label{Dirac equation 3}
\end{eqnarray}
Then $\Phi(r,\theta,\phi)$ can be separated into radial and angular parts by defining \cite{Cho:2004wj,Bjorken:1964xl}
\begin{eqnarray}
\Phi^{(\pm)}(r,\theta,\phi)=\left( \begin{array}{c}
    \text{i}G^{(\pm)}(r)\varphi^{(\pm)}_{jm}(\theta,\phi) \\
    F^{(\pm)}(r)\varphi^{(\mp)}_{jm}(\theta,\phi)
\end{array} \right). \label{Phi}
\end{eqnarray}
The angular parts $\varphi^{(\pm)}_{jm}(\theta,\phi)$ are eigenfunctions of the operator $\mathrm{k}=\vec{\sigma}\cdot\vec{L}+\mathbf{1}$: $\mathrm{k} \varphi^{(\pm)}_{jm}(\theta,\phi)=\pm(j+1/2)  \varphi^{(\pm)}_{jm}(\theta,\phi)$. Their explicit forms are given by
{
\begin{eqnarray}
\varphi^{(+)}_{jm}&=&\left( \begin{array}{c}
    \sqrt{\frac{j+m}{2j}}Y^{m-1/2}_{j-1/2} \\
    \sqrt{\frac{j-m}{2j}}Y^{m+1/2}_{j-1/2}
\end{array} \right),\\
\varphi^{(-)}_{jm}&=&\left(\begin{array}{c}
     \sqrt{\frac{j+1-m}{2j+2}}Y^{m-1/2}_{j+1/2}\\
     -\sqrt{\frac{j+1+m}{2j+2}}Y^{m+1/2}_{j+1/2}
\end{array}\right).\label{varphi}
\end{eqnarray}
}
Here $j$ is the total angular momentum quantum number and $j\in\{\frac{1}{2},\frac{3}{2},\frac{5}{2},\cdots\}$. $m$ is the magnetic quantum number and $m=-j, -j+1, \cdots, j$. $Y^{m\pm1/2}_{j\pm1/2}$ are the spheric harmonics functions satisfying $\int^{2\pi}_{\phi=0}\int^{\pi}_{\theta=0}Y^{n}_{l}Y^{n'*}_{l'}\sin\theta d\theta d\phi=\delta_{ll'}\delta_{nn'}$. $\Phi^{(\pm)}$ are eigenfunctions of the spin-orbit operator $K= \beta (\vec{\sigma}\cdot\vec{L}+\mathbf{1})$: $K \Phi^{(\pm)}=\kappa_{(\pm)} \Phi^{(\pm)}$ ($\kappa_{(\pm)}=\pm(j+1/2)$) \cite{Thaller:1992xl}. For the $(+)$ case, it will be convenient to make a transformation of variables \cite{Cho:2004wj}:
\begin{eqnarray}
\left(
\begin{array}{c}
\hat{F}^{(+)}\\ \hat{G}^{(+)} \end{array} \right) =\left(
\begin{array}{cc}
{\rm sin}\frac{\theta_{(+)}}{2}  &   {\rm cos}\frac{\theta_{(+)}}{2}\\
{\rm cos}\frac{\theta_{(+)}}{2}  &  -{\rm sin}\frac{\theta_{(+)}}{2}
\end{array}\right)
\left(
\begin{array}{c}
F^{(+)}\\G^{(+)}
\end{array}
\right),\label{transformation+}
\end{eqnarray}
 where $\theta_{(+)}=\tan^{-1}(mr/\kappa_{(+)})$. Similarly, for the $(-)$ case, we can make a transformation
\begin{eqnarray}
\left(
\begin{array}{c}
\hat{F}^{(-)}\\ \hat{G}^{(-)} \end{array} \right) =\left(
\begin{array}{cc}
{\rm cos}\frac{\theta_{(-)}}{2} & -{\rm sin}\frac{\theta_{(-)}}{2}\\
{\rm sin}\frac{\theta_{(-)}}{2} &  {\rm cos}\frac{\theta_{(-)}}{2}
\end{array}\right)
\left(
\begin{array}{c}
F^{(-)}\\G^{(-)}
\end{array}
\right)\label{transformation-}
\end{eqnarray}
with $\theta_{(-)}=\tan^{-1}(mr/\kappa_{(-)})$. Substituting Eqs. (\ref{Phi}), (\ref{transformation+}), and (\ref{transformation-}) into Eq. (\ref{Dirac equation 3}) and making a transformation of variable $r$:
\begin{eqnarray}
{\hat{r}_{*}}=r+2M\ln\big(\frac{r}{2M}-1\big)+\frac{1}{2\omega}\tan^{-1}\left(\frac{mr}{\kappa}\right),
\end{eqnarray}
we can get the following radial equations of $\hat{F}$ and $\hat{G}$:
\begin{eqnarray}
\left(-\frac{d^{2}}{d{\hat{r}_{*}}^{2}}+V_{1}\right)\hat{F}
&=&\omega^{2}\hat{F},\label{EF}\\
\left(-\frac{d^{2}}{d{\hat{r}_{*}}^{2}}+V_{2}\right)\hat{G}
&=&\omega^{2}\hat{G}.\label{EG}
\end{eqnarray}
The effective potentials $V_{1}$ and $V_{2}$ are given by
\begin{eqnarray}
V_{1,2}=\pm\frac{dW}{d{\hat{r}_{*}}}+W^{2},\label{V12}
\end{eqnarray}
where
\begin{eqnarray}
W=\frac{\sqrt{N}(\kappa^{2}+m^{2}r^{2})^{3/2}}{r(\kappa^{2}+m^{2}r^{2}+m\kappa N/2\omega)}. \label{W}
\end{eqnarray}
Here the cases of $(+)$ and $(-)$ have been combined, so $\kappa$ covers all positive and negative integers. As in the scalar field case, we can get resonance states of the Dirac field by solving these Schr\"odinger-like equations (\ref{EF}) and (\ref{EG}). There are two different potentials $V_{1}$ and $V_{2}$, but in fact, they are both derived from a same superpotential $W$ according to Eq. (\ref{V12}). The two potentials are called supersymmetric partners, which will result in the same spectra of quasinormal modes. The quasinormal modes and stationary resonant states are related as shown in Ref.~ \cite{PhysRevLett.109.081102}. Hence, Eqs. (\ref{EF}) and (\ref{EG}) should also give the same spectra of stationary resonance states.

\section{Stationary resonance states of Dirac field}
\label{section3}
In this section, we will look for the stationary resonant states of the Dirac field using the Schr\"odinger-like equation (\ref{EF}). Substituting Eq. (\ref{W}) into Eq. (\ref{V12}), the effective potential $V_1$ can be written as \cite{Cho:2004wj}
\begin{eqnarray}
&&V_{1}(r,\kappa,m,\omega)\nonumber\\
&=&\Big[rN^{1/2}(r)(\kappa^{2}+m^{2}r^{2})^{3/2}\nonumber \\
  &&+(r-1)(\kappa^{2}+m^{2}r^{2})+3m^{2}r^{3}N(r)\Big]\nonumber\\
  &&\times\frac{N^{1/2}(r)(\kappa^{2}+m^{2}r^{2})^{3/2}}{r^{3}(\kappa^{2}+m^{2}r^{2}+m\kappa N(r)/2\omega)^{2}}\nonumber \\
  &&-\frac{N^{3/2}(r)(\kappa^{2}+m^{2}r^{2})^{5/2}}{r^{3}(\kappa^{2}+m^{2}r^{2}+m\kappa N(r)/2\omega)^{3}}\nonumber\\
  &&\times\left[2r(\kappa^{2}+m^{2}r^{2})+2m^{2}r^{3}+m\kappa(r-1)/\omega\right].
\label{massiveV}
\end{eqnarray}
Obviously, the potential depends not only on the mass of field $m$, but also on the parameter $\kappa$ related to the spin-orbit interaction and the frequency $\omega$. The behavior of the effective potential have been analyzed in detail in Refs. \cite{PhysRevD.68.024003} and \cite{Cho:2004wj}. As analyzed, for small values of $m$, the potential behaves as a barrier and its asymptotic value when $r$ approaches infinity is
\begin{eqnarray}
V_{1}(r\rightarrow\infty)=m^2.
\end{eqnarray}
On the other hand, for fixed $m$ and $\omega$, the peak of the potential will increase with $|\kappa|$. Taking the limit $|\kappa|\rightarrow\infty$, we can get
\begin{eqnarray}
V_{1}(|\kappa|\rightarrow\infty)\thickapprox\frac{N(r)\kappa^2}{r^2},
\end{eqnarray}
which has the maximum value $\frac{\kappa^2}{27M^2}$ at $r=3M$. Particularly, if we fix $m$, $\omega$, and the value of $|\kappa|$, the peak of the potential will be higher in the case with negative $\kappa$. This behavior leads to interesting results which we will show in the following sections.  Finally, we consider the dependence of the potential on the frequency $\omega$. We will see later that all the values of frequency $\omega$ we concern are within a small region near the mass $m$.  Thus the frequency $\omega$ does not change the general behavior of the potential within our consideration in this paper.

\begin{figure}[htbp]
\begin{center}
\includegraphics[width=3in]{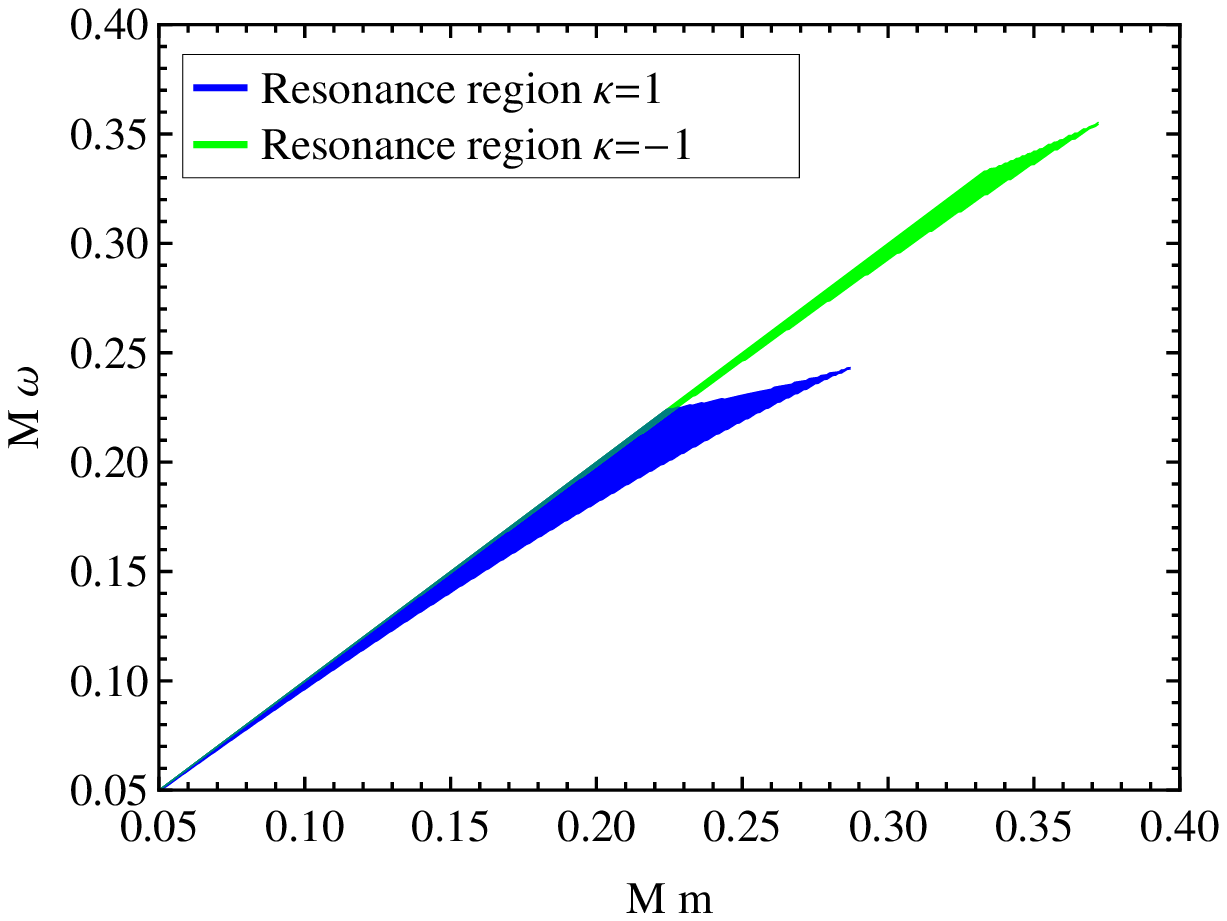}
\includegraphics[width=3in]{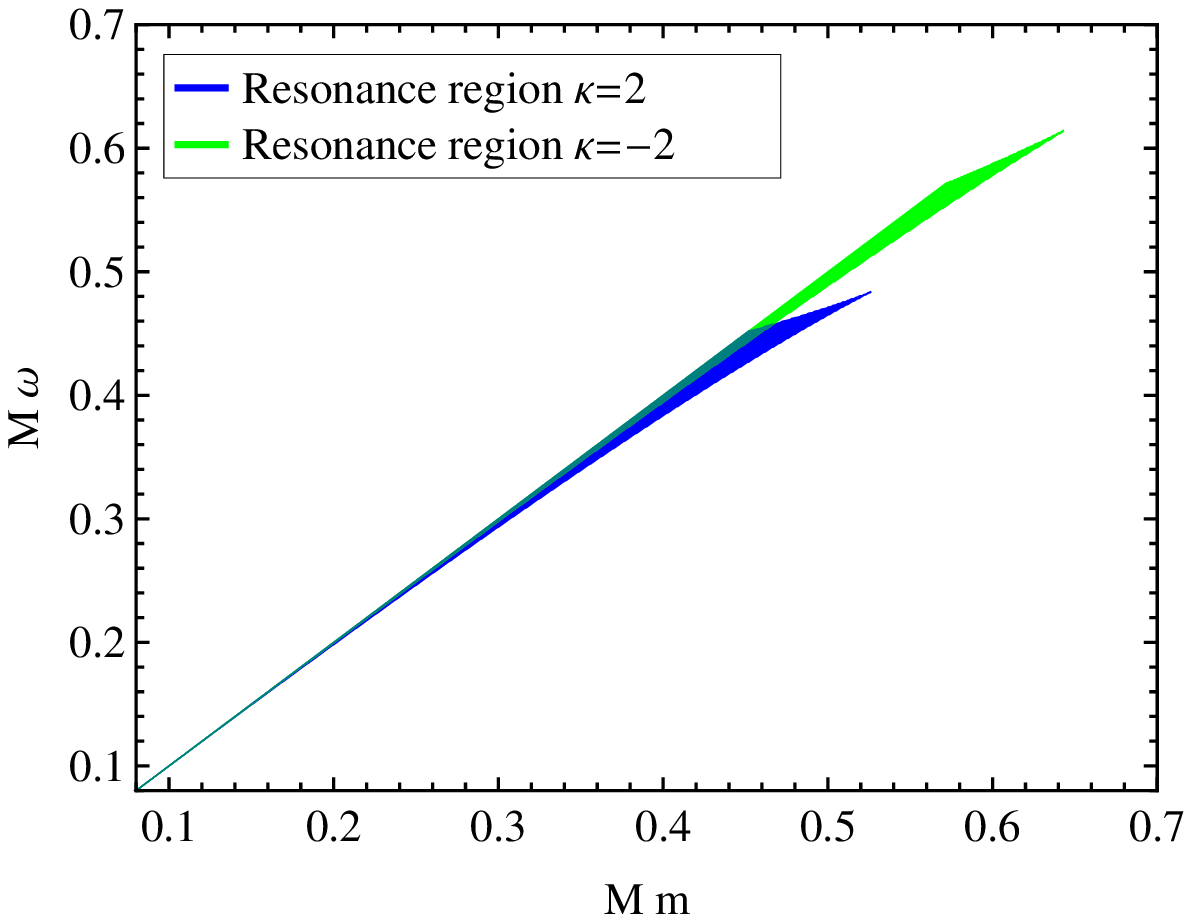}
\includegraphics[width=3in]{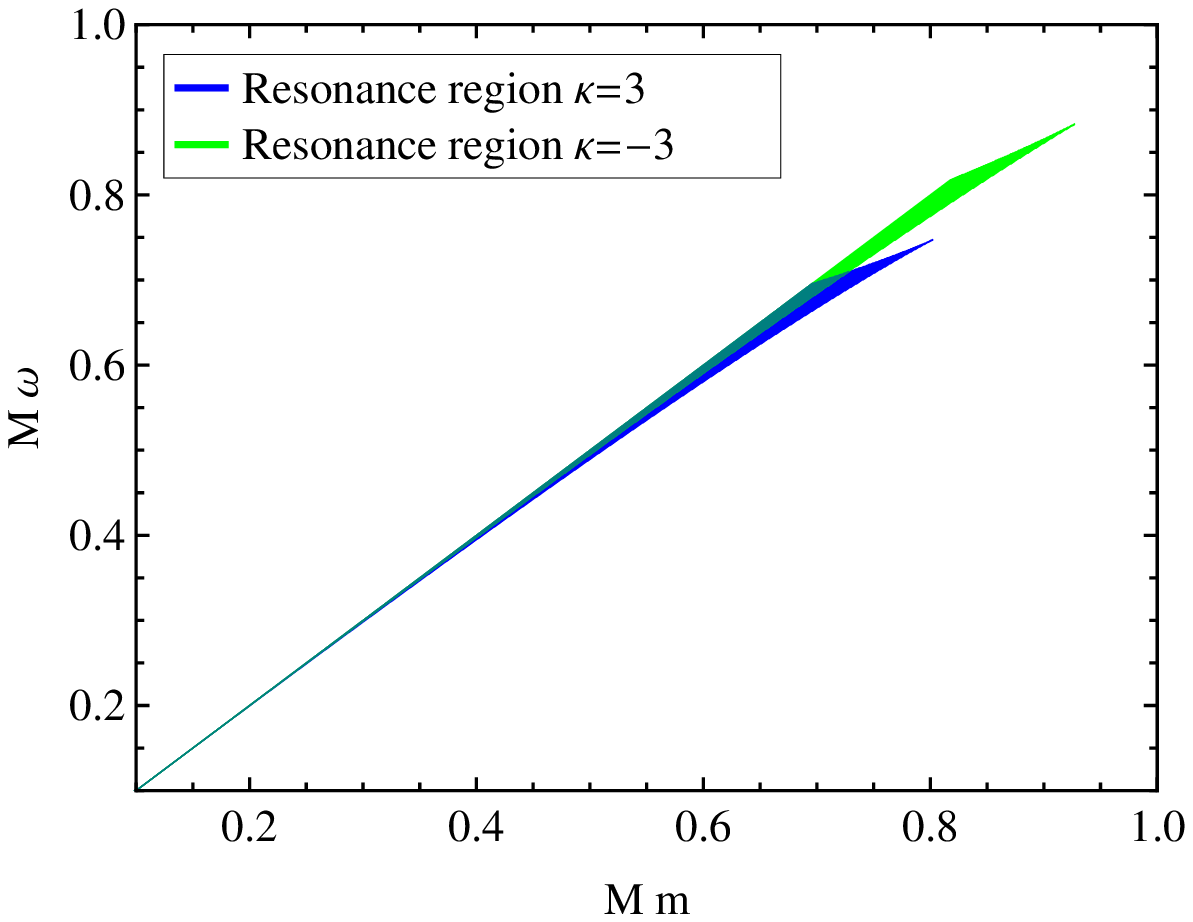}
\caption{{Resonance band for positive $\omega$ and different values of $\kappa$.}}
\label{band}
\end{center}
\end{figure}

As in the scalar case \cite{PhysRevD.84.083008}, it is also possible to have resonance states when $\omega$ lies in the ``resonance band", i.e., $V_1^{min}<\omega^2<\{V_1^{max}, m^2\}$ (The states with $\omega$ lying outside the ``resonance band" are not within our consideration because initial data constructed from them typically have much shorter lifetime). To find the ``resonance band",  we need to get the extremities of the potential. But it is difficult to solve the equation $\partial_{r}V_{1}(r)=0$ analytically. And noting that the effective potential also depends on $\omega$. Thus we perform a numerical analysis and show the ``resonance band" for some values of $\kappa$ in Fig. \ref{band}. Note that $\kappa$ goes over all the positive and negative integers, and the frequency $\omega$ covers all real number. As discussed in the previous section, there are two solutions for Eq.~(\ref{Dirac equation 3}), labeled as $(+)$ and $(-)$ respectively, which are related to spin-orbit interaction. On the other hand, the negative $\omega$ will lead to negative energy related to antiparticle states. However, it is easy to find from Eq. (19) that if we change both the signs of $\kappa$ and $\omega$, the potential $V_1$ will remain unchanged. Thus we just need to discuss two cases: $M\omega \kappa>0$ and $M\omega \kappa<0$.

Fig. \ref{band} shows the resonance band for $\kappa=\pm1$, $\kappa=\pm2$, and $\kappa=\pm3$, where we have chosen $\omega>0$. It is shown that the cutoff of $Mm$ increases with the growth of $|\kappa|$. For the same value of $|\kappa|$, the cutoff of $Mm$ is larger for the case with negative $\kappa$. And the resonance region of $\omega$ becomes narrower and narrower when $Mm$ approaches 0.

\begin{figure}[htbp]
\begin{center}
\includegraphics[width=3in]{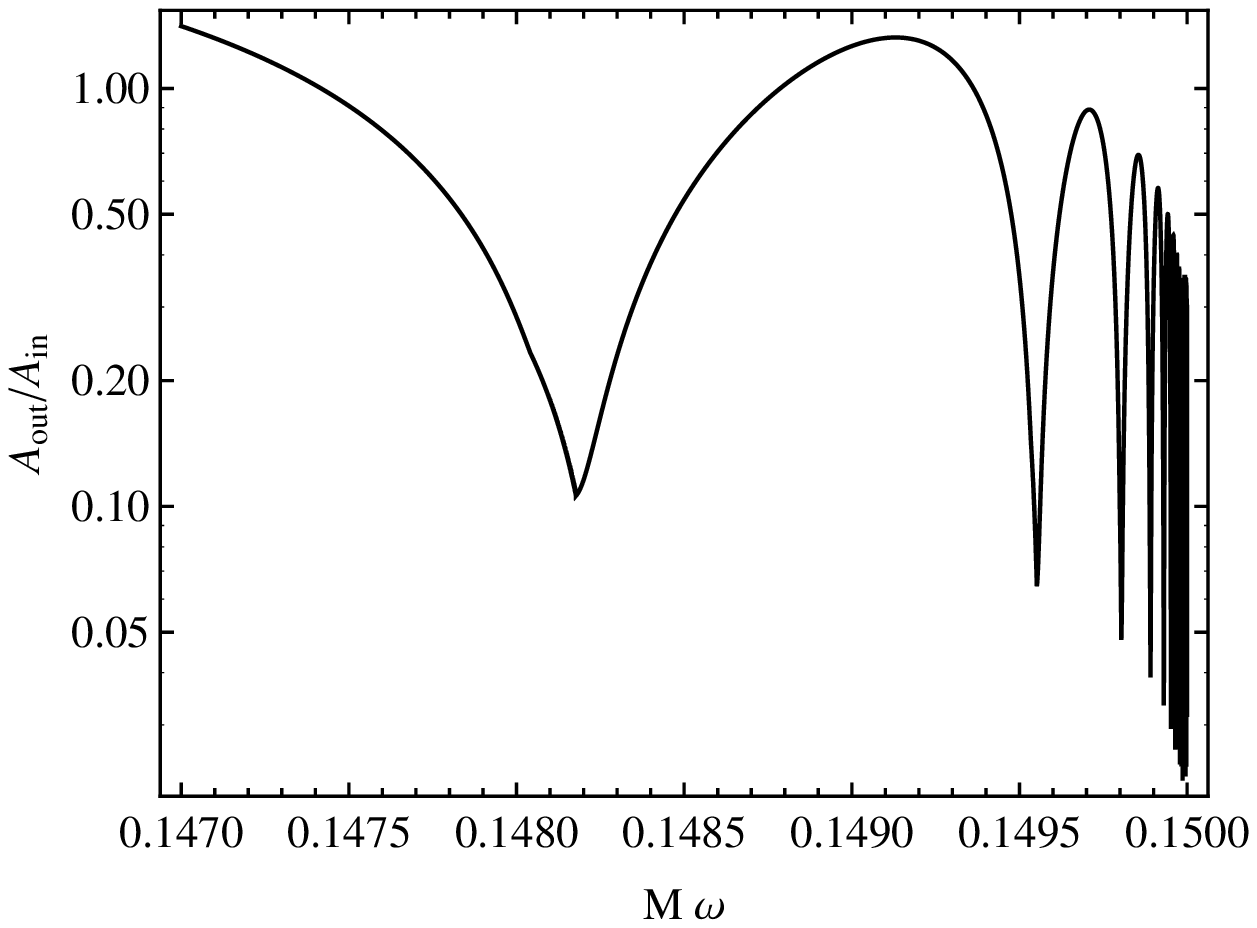}
\includegraphics[width=3.05in]{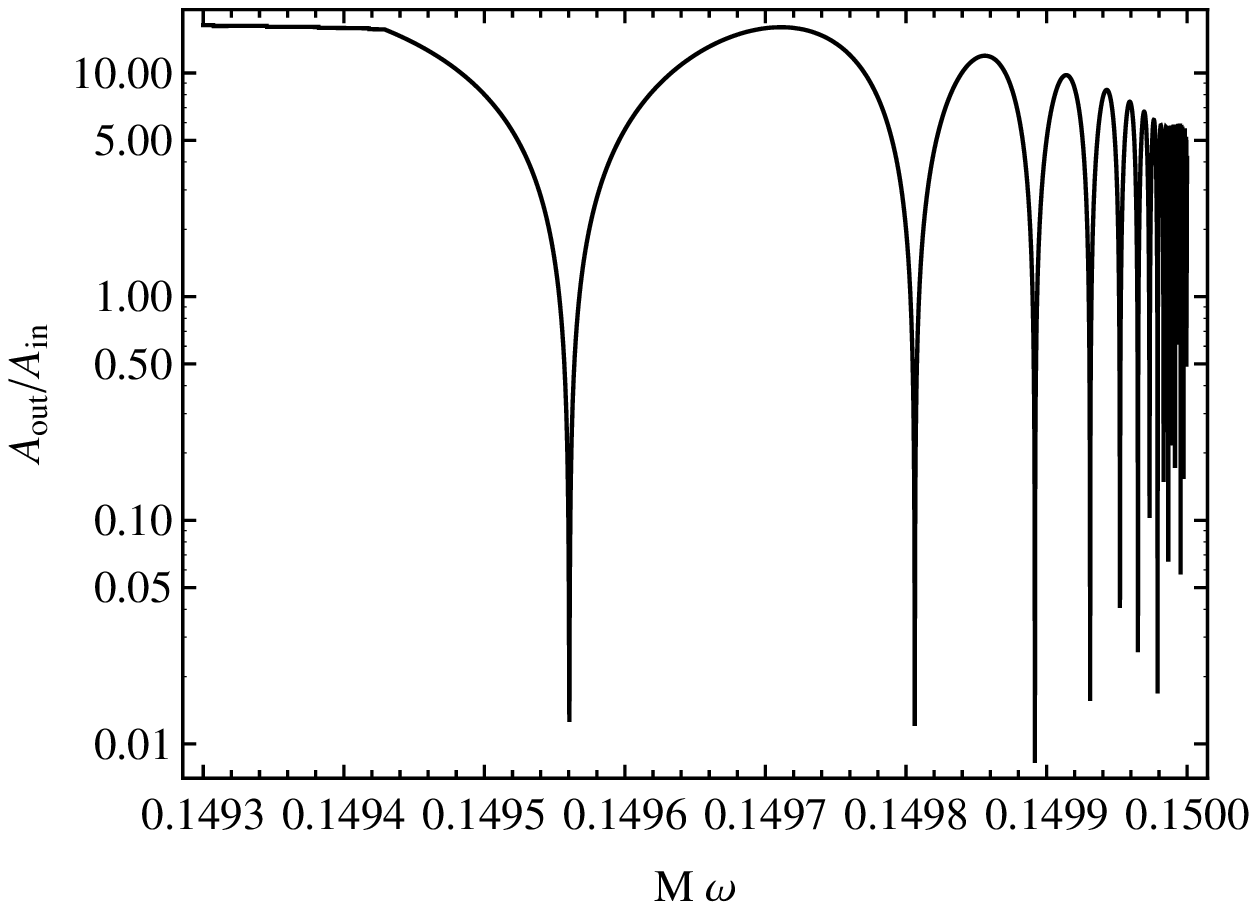}
\caption{Resonant frequencies for $Mm=0.15$. $\kappa$ is taken to be $+1$ (upper plot) and $-1$ (lower plot). $A_{out}$ and $A_{in}$ are the maximum values of the amplitude of $\hat{F}$ for $\hat{r}_{*}/M<0$ and $\hat{r}_{*}/M>10$ respectively.}
\label{Resonant}
\end{center}
\end{figure}

Now, we can solve Eq. (\ref{EF}) (or Eq. (\ref{EG})) numerically to find the resonance spectra. Following the analysis in Ref. \cite{PhysRevD.84.083008}, we choose appropriate boundary conditions to ensure that $\hat{F}$ is real and decays exponentially at large $r$. As in the scalar case, for some discrete frequencies $\omega$, the amplitude for $\hat{r}_{*}/M<0$ is much smaller than the amplitude in the potential well. These discrete frequencies are called resonant frequencies. Fig.~\ref{Resonant} shows the ratio $A_{out}/A_{in}$, where $A_{out}$ and $A_{in}$ are the maximum values of the amplitude of $\hat{F}$ for $\hat{r}_{*}/M<0$ and $\hat{r}_{*}/M>10$, respectively. It is easy to find that the resonant frequencies become closer and closer to each other when $\omega$ approaches $m$. We find an interesting result that the resonant spectra of the cases with the same $m$ and $|\kappa|$ are almost the same, except for the first resonant frequency of the one with positive $\kappa$ (see Tab. \ref{Resonantpoint}).

\begin{table}[htbp]
\caption{Resonant frequencies $M\omega$ for different masses of the Dirac field $Mm$. We can find that for the same $Mm$, the second resonant frequency of the case $\kappa=1$ is very close to the first resonant frequency of the case $\kappa=-1$. Thence, the two cases almost have the same resonant spectra. \label{tab1}}
\begin{center}
\begin{tabular}{l l c c c c c}
\hline \hline
$Mm\;\;\;\;$    &
${}\;\;\;\;\;\;\;\;\;\;\;$ &
$\;\;\;\;\;\;\;\;0.10\;\;\;\;\;\;\;$ &
$\;\;\;\;\;\;\;\;0.15\;\;\;\;\;\;\;$ &
$\;\;\;\;\;\;\;\;0.20\;\;\;\;\;\;\;$\\
\hline
 $n=1$  & $\kappa=+1$ &  0.0995079 & 0.1481772  & 0.1950961 \\
${}$    & $\kappa=-1$&  0.0998728 & 0.1495600  & 0.1989210 \\
$n=2$   & $\kappa=+1$ &  0.0998750 & 0.1495518  & 0.1988426 \\
${}$    & $\kappa=-1$&  0.0999437 & 0.1498060  & 0.1995276 \\
$n=3$   & $\kappa=+1$ &  0.0999443 & 0.1498038  & 0.1995059 \\
${}$    & $\kappa=-1$&  0.0999684 & 0.1498920  & 0.1997373 \\
$n=4$   & $\kappa=+1$ &  0.0999687 & 0.1498907  & 0.1997284 \\
${}$    & $\kappa=-1$&  0.0999798 & 0.1499310  & 0.1998333 \\
$n=5$   & $\kappa=+1$ &  0.0999800 & 0.1499305  & 0.1998287 \\
${}$    & $\kappa=-1$&  0.0999860 & 0.1499520  & 0.1998849 \\
\hline\hline\label{Resonantpoint}
\end{tabular}
\end{center}
\end{table}


\section{Numerical evolution of the Dirac resonance configurations}
\label{section4}
\subsection{Initial data}
\label{initialdate}
{ As discussed in Refs.~\cite{PhysRevD.84.083008,PhysRevLett.109.081102}, the stationary solutions are indeed nonphysical, because they
have divergent energy due to their oscillatory behavior close to the horizon. And the ingoing and outgoing modes of these solutions have the same
amplitude close to the horizon, which implies that any Dirac particles falling into the black hole are compensated by the Dirac particles escaping from the horizon. This is also nonphysical. However, we can construct physical configurations, the so-called ``pseudo-stationary configurations",
which are very close to the stationary solutions. It is done by truncating the rapid oscillatory parts near the horizon of the stationary solutions and set them to zero by hand. Then the configurations we obtain have regular behavior close to the horizon. They can be seen as a combination of the
stationary solutions and perturbations around the horizon (for more specific discussions, see Ref.~\cite{PhysRevD.84.083008}).}


Substituting Eqs. (\ref{Phi}) and (\ref{varphi}) into Eq. (\ref{Dirac equation 3}), we obtain two coupled first-order equations (the $(+)$ and $(-)$ cases have been combined and $\kappa$ covers all positive and negative integers as mentioned before)
\begin{eqnarray}
(m-\omega N^{-\frac{1}{2}}(r))G&=&\left(N^{\frac{1}{2}}(r)\partial_{r}+\frac{\kappa}{r}\right)F,\label{EFG1}\\
(m+\omega N^{-\frac{1}{2}}(r))F&=&\left(N^{\frac{1}{2}}(r)\partial_{r}-\frac{\kappa}{r}\right)G.\label{EFG2}
\end{eqnarray}
Note that when $r\rightarrow+\infty$, the above equations will become
\begin{eqnarray}
(m-\omega)G=\partial_{r}F,\qquad (m+\omega)F=\partial_{r}G,
\end{eqnarray}
which can be rewritten as
\begin{eqnarray}
\partial^{2}_{r}G=(m^2-\omega^2)G,\qquad F=\frac{1}{m+\omega}\partial_{r}G.
\end{eqnarray}
Obviously, when $r\rightarrow+\infty$, $G\rightarrow\text{e}^{-k_{\omega}r}$, and $F\rightarrow-\sqrt{\frac{m-\omega}{m+\omega}}\text{e}^{-k_{\omega}r}$ with $k_{\omega}=\sqrt{m^2-\omega^2}$. In order to get the initial data we need, i.e., the pseudo-stationary initial data, we first solve Eqs. (\ref{EFG1}) and (\ref{EFG2}) in the region $[R_{in}=2M+\epsilon,R_{out}]$  with $\epsilon\lesssim0.05M$. Then rescale the solution so that {$|G(R_{in})|=1$}. Finally we set $G$ and $F$ to 0 for $r<R_{in}$. Note that if we change both the signs of $\omega$ and $\kappa$ in Eqs. (\ref{EFG1}) and (\ref{EFG2}) and exchange $F$ and $G$, the solutions will remain unchanged. So we just need to consider two cases as mentioned before: $M\omega\kappa>0$ and $M\omega\kappa<0$.

\subsection{Numerical evolution}

There are two kinds of initial data when $\omega$ lies in the resonance band: pseudo-stationary configurations constructed from resonant stationary states and configurations constructed from non-resonant stationary states. We will concentrate on the first kind, because those configurations can last longer in the potential well. As in the scalar case, we perform the numerical evolution in the ingoing Eddington-Finkelstein coordinates which can cover the region in the event horizon, so that we do not need to impose the left boundary conditions. In these coordinates,
\begin{eqnarray}
  ds^2 &=& -(1-{\Delta})d\bar{t}^2+2{\Delta}d\bar{t}dr\nonumber \\
       &&  +(1+{\Delta})dr^2+r^2\left(d\theta^2+\sin^2\theta d\phi^2\right),
\end{eqnarray}
where ${\Delta}=\frac{2M}{r}$ and $\bar{t}:=t+2M\ln(r/2M-1)$. This coordinate system can cover the region $r\in(0,\infty)$ and for a fixed radial coordinate $r$, the Schwarzschild and Eddington-Finkelstein times can be considered equivalent: $\Delta t=\Delta \bar{t}$. So we drop the bar on Eddington-Finkelstein time in the following discussion. To be compatible with the Eddington-Finkelstein metric, we introduce the vierbein
\begin{eqnarray}
{e_{a}}^{\mu} = \left(
\begin{array}{cccc}
 \sqrt{1+\Delta}&-\frac{{\Delta}}{\sqrt{1+\Delta}}&0&0\\
 0 &\frac{\sin\theta\cos\phi}{\sqrt{1+\Delta}}  & \frac{\cos\theta\cos\phi}{r} & -\frac{\csc\theta\sin\phi}{r} \\
 0 &\frac{\sin\theta\sin\phi}{\sqrt{1+\Delta}}  & \frac{\cos\theta\sin\phi}{r} & \frac{\csc\theta\cos\phi}{r}   \\
 0 &\frac{\cos\theta}{\sqrt{1+\Delta}}          & -\frac{\sin\theta}{r}        & 0                          \\
\end{array}\right).~~~~
\end{eqnarray}
Then Eq.~(\ref{Dirac equation 1}) can be written as
\begin{eqnarray}
&&\gamma^{0}\sqrt{1+\Delta}\partial_{t}\Psi
 +\frac{\tilde{\gamma}-\Delta\gamma^{0}}{r(1+\Delta)^{\frac{1}{4}}}
   \partial_{r}\left(\frac{r}{(1+\Delta)^{\frac{1}{4}}}\Psi\right)\nonumber\\
&&+\frac{\Delta\gamma^{0}}{2r\sqrt{1+\Delta}}\Psi-\frac{\tilde{\gamma}}{r}\left(\vec{\Sigma}\cdot\vec{L}+1\right)\Psi+m\Psi=0.
\label{Dirac equation 4}
\end{eqnarray}
Introducing a new decomposition
{
\begin{eqnarray}
\Psi(t,r,\theta,\phi)&=&\frac{(1+\Delta)^{\frac{1}{4}}}{r}\tilde{\Phi}(t,r,\theta,\phi),\\ \tilde{\Phi}(t,r,\theta,\phi)&=&\left( \begin{array}{c}
    \text{i}\tilde{G}^{(\pm)}(t,r)\varphi^{(\pm)}_{jm}(\theta,\phi) \\
    \tilde{F}^{(\pm)}(t,r)\varphi^{(\mp)}_{jm}(\theta,\phi)
\end{array} \right), \label{Phi2}
\end{eqnarray}
}
Eq.~(\ref{Dirac equation 4}) reduces to the following two coupled first-order equations:
\begin{eqnarray}
 &&\left(\sqrt{1+{\Delta}}\partial_{t}
  -\frac{{\Delta}}{\sqrt{1+{\Delta}}}\partial_{r}
  +\frac{{\Delta}}{2r\sqrt{1+{\Delta}}}+\text{i}m\right)\tilde{G}\nonumber\\
 &=&\text{i}\left(\frac{1}{\sqrt{1+{\Delta}}}\partial_{r}
    +\frac{\kappa}{r}\right)\tilde{F},~~~~~~~\label{EF3}\\
 &&\left(\sqrt{1+{\Delta}}\partial_{t}
  -\frac{{\Delta}}{\sqrt{1+{\Delta}}}\partial_{r}
  +\frac{{\Delta}}{2r\sqrt{1+{\Delta}}}-\text{i}m\right)\tilde{F}\nonumber\\
 &=&\text{i}\left(-\frac{1}{\sqrt{1+{\Delta}}}\partial_{r}
  +\frac{\kappa}{r}\right)\tilde{G}.~~~~~~~\label{EF4}
\end{eqnarray}
Here we have combined the cases $(+)$ and $(-)$ and $\kappa=\pm 1,\pm 2,\pm 3,\cdots$. Obviously, $\tilde{G}(t,r)$ and $\tilde{F}(t,r)$ should be complex functions and can be decomposed into real parts and imaginary parts: $\tilde{G}(t,r)=G_{R}(t,r)+\text{i}G_{I}(t,r)$ and $\tilde{F}(t,r)=F_{R}(t,r)+\text{i}F_{I}(t,r)$. Then we can obtain four real equations:
\begin{eqnarray}
\hat{A}G_{R}-mG_{I}&=&-\hat{B}_{+}F_{I},~~ \hat{A}G_{I}+mG_{R}=\hat{B}_{+}F_{R},\label{AG}\\
\hat{A}F_{R}+mF_{I}&=&\hat{B}_{-}G_{I},~~~~~ \hat{A}F_{I}-mF_{R}=-\hat{B}_{-}G_{R},~~~\label{AF}
\end{eqnarray}
where $\hat{A}=\sqrt{1+{\Delta}}\partial_{t}-\frac{{\Delta}}{\sqrt{1+{\Delta}}}\partial_{r}+\frac{{\Delta}}{2r\sqrt{1+{\Delta}}}$ and $\hat{B}_{\pm}=\frac{1}{\sqrt{1+{\Delta}}}\partial_{r}\pm\frac{\kappa}{r}$.

\begin{figure}[htbp]
\begin{center}
~~~
\includegraphics[width=3in]{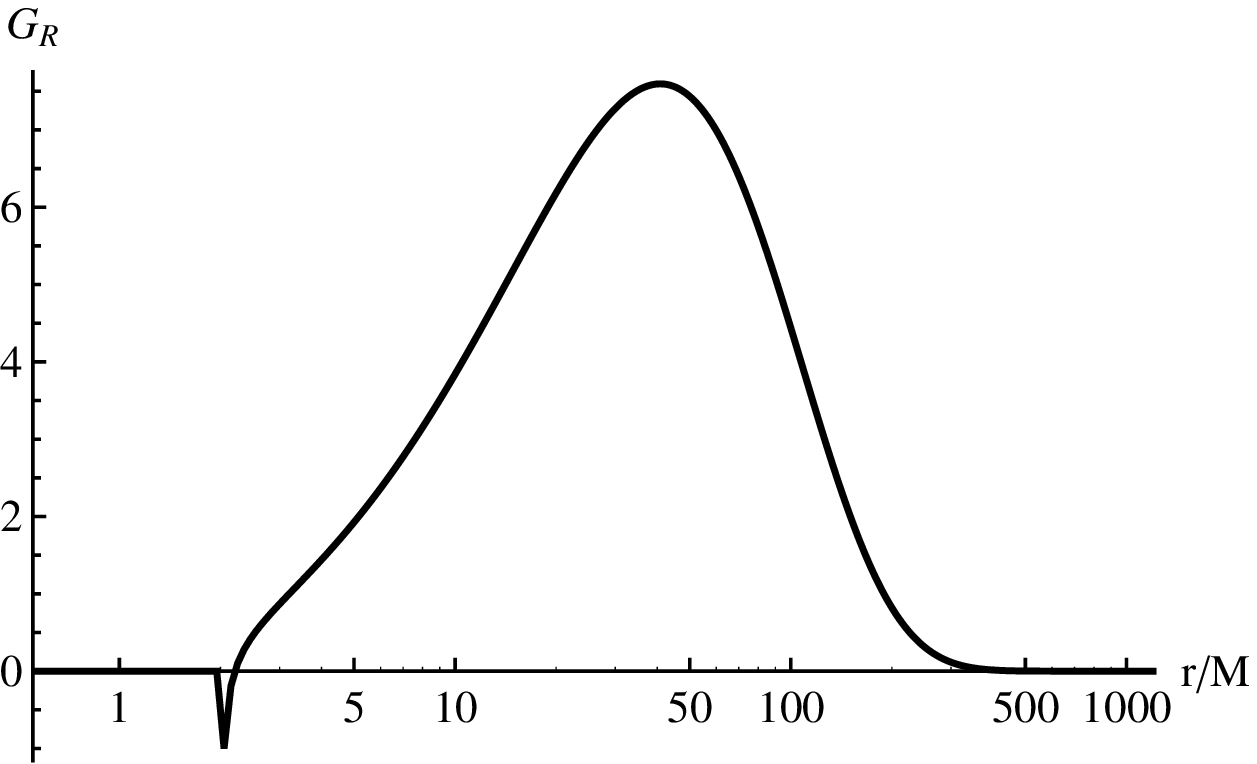}
\includegraphics[width=3.19in]{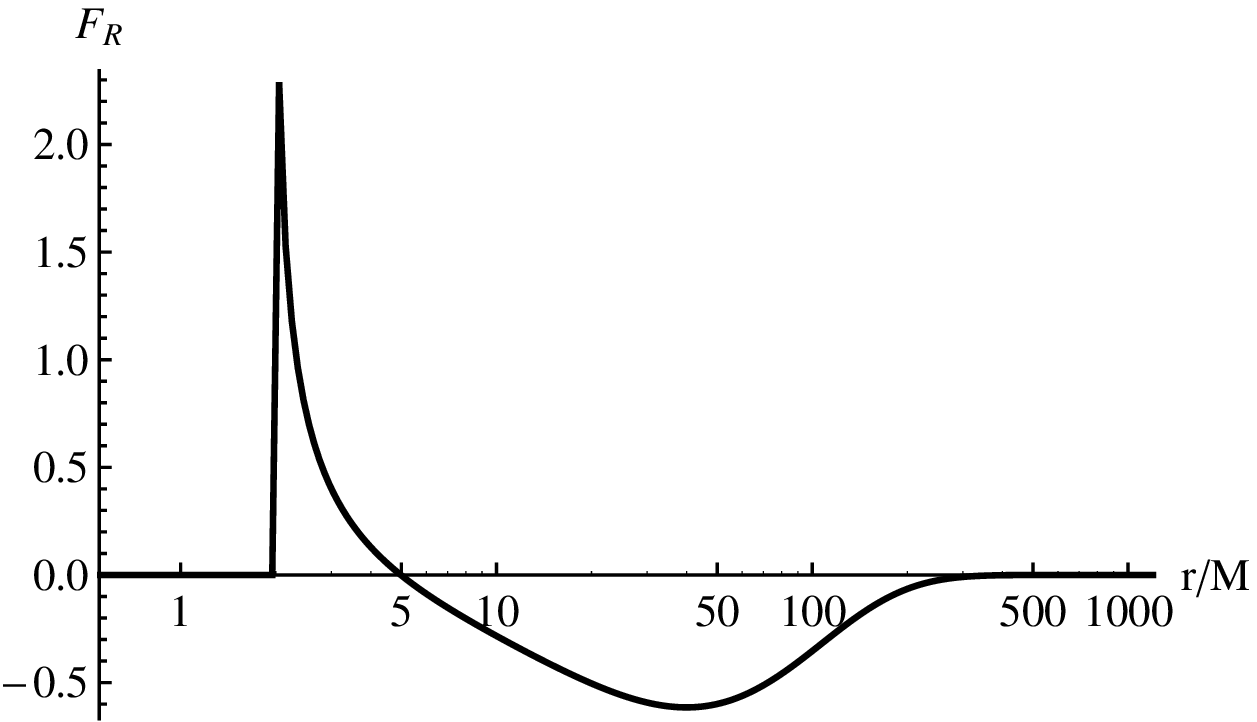}
\caption{{Initial configurations of pseudo-resonant state with $Mm=0.15$, $\kappa=1$, and $n=1$.}}
\label{initial}
\end{center}
\end{figure}

To solve Eqs. (\ref{AG}) and (\ref{AF}), we need initial data and right boundary conditions for $G_R$, $G_I$, $F_R$, and $F_I$. We suppose $\tilde{G}$ and $\tilde{F}$ are both real at $t=0$ (this can be done by choosing a special phase without loss of generality), and their real parts equal to the pseudo-stationary states as discussed before. {A typical example of initial configurations is shown in Fig. \ref{initial}.} Then we impose the maximally dissipative boundary conditions \cite{Degollado:2009rw}: $G_R(R_{out})-F_I(R_{out})=0$ and $G_I(R_{out})+F_R(R_{out})=0$. Finally with these initial data and boundary conditions, we solve Eqs. (\ref{AG}) and (\ref{AF}) numerically using fourth-order finite differences in space, and evolving in time using an explicit embedded Runge-Kutta Prince-Dormand (8, 9) method.

Similar to that in Ref. \cite{PhysRevD.84.083008}, we also define the energy of the Dirac field
\begin{eqnarray}
E=\int_{\Sigma}k_{0}^{\mu}T_{\mu\nu}n^{\nu}\sqrt{\gamma}d\Sigma,
\end{eqnarray}
where $k_0^{\mu}=\partial_t$ is a Killing vector field, $n^{\nu}$ is the normal vector of the 3-dimensional space-like hypersurface $\Sigma=\{t=\text{constant}\}$, and $\sqrt{\gamma}d\Sigma$ is the volume element on $\Sigma$. For more details, refer to Ref.~\cite{PhysRevD.84.083008}. $T_{\mu\nu}$ is the energy-momentum tensor of the Dirac field. Note that the Lagrangian density of the Dirac field is
\begin{eqnarray}
\mathfrak{L}_{\text{Dirac}}=-\frac{1}{2}\bar{\Psi}\Gamma^{\mu}\overleftrightarrow{\mathfrak{D}}_{\mu}\Psi-m\bar{\Psi}\Psi.
\end{eqnarray}
Here $\bar{\Psi}=\text{i}\Psi^{\dag}\gamma^{0}$ and $\bar{\Psi}\Gamma^{\mu}\overleftrightarrow{\mathfrak{D}}_{\mu}\Psi=\bar{\Psi}\Gamma^{\mu}\mathfrak{D}_{\mu}\Psi-\bar{\Psi}\overleftarrow{\mathfrak{D}}_{\mu}\Gamma^{\mu}\Psi$ with $\bar{\Psi}\overleftarrow{\mathfrak{D}}_{\mu}=\partial_{\mu}\bar{\Psi}-\bar{\Psi}\Omega_{\mu}$. Hence the energy-momentum tensor $T_{\mu\nu}$ can be written as \cite{Forger:2003ut}
\begin{eqnarray}
T_{\mu\nu}=\frac{1}{4}\left(\bar{\Psi}\Gamma_{\mu}\overleftrightarrow{\mathfrak{D}}_{\nu}\Psi+\bar{\Psi}\Gamma_{\nu}\overleftrightarrow{\mathfrak{D}}_{\mu}\Psi\right)-g_{\mu\nu}\mathfrak{L}_{\text{Dirac}}
\end{eqnarray}
and the energy of a state with specific ``spin-orbit interaction quantum number" $\kappa$ and magnetic quantum number $m$ (we should distinguish the ``$m$" here with the mass of the Dirac field) $E_{\kappa m}$ is given by
\begin{eqnarray}
E_{\kappa m}=\int^{\infty}_{2M}\rho_{E}dr,
\end{eqnarray}
where
\begin{eqnarray}
    \rho_{E}&=&\left(1+{\Delta}\right)
               \Big[(\partial_{t}G_{R}G_{I}-G_{R}\partial_{t}G_{I}
                     +\partial_{t}F_{R}F_{I}-F_{R}\partial_{t}F_{I})\nonumber\\
            &+&\frac{{\Delta}}{2}
               \big(F_{R}\partial_{t}G_{R}+F_{I}\partial_{t}G_{I}
                      -G_{R}\partial_{t}F_{R}-G_{I}\partial_{t}F_{I}\big)
               \Big]\nonumber\\
            &-&\frac{{\Delta}}{2}
               \Big(\partial_{r}G_{R}G_{I}-G_{R}\partial_{r}G_{I}
                   +\partial_{r}F_{R}F_{I}-F_{R}\partial_{r}F_{I}\Big)\nonumber\\
            &-&\frac{\Delta^2\!\!}{2}
               \Big(F_{R}\partial_{r}G_{R}+F_{I}\partial_{r}G_{I}-G_{R}\partial_{r}F_{R}-G_{I}\partial_{r}F_{I}\Big).~~~
\end{eqnarray}

\begin{figure}[htbp]
\begin{center}
\includegraphics[width=3in]{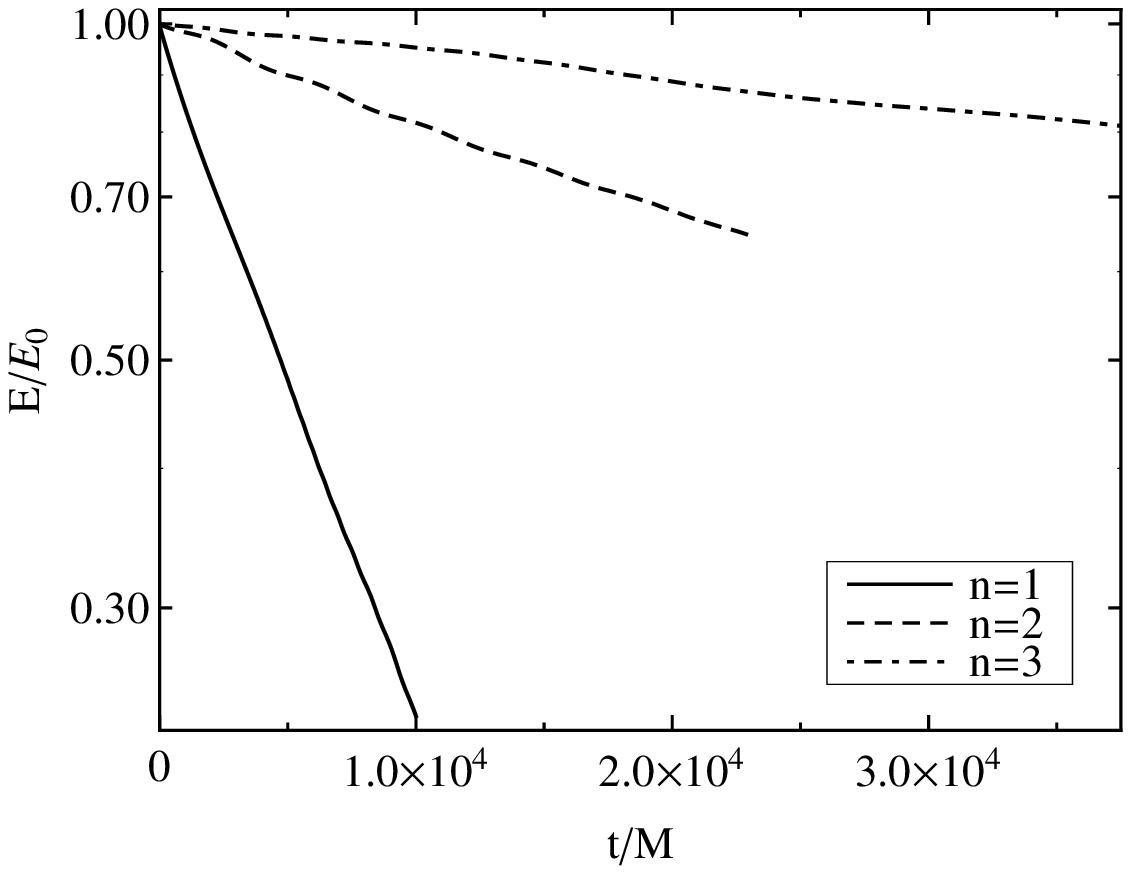}
\includegraphics[width=3in]{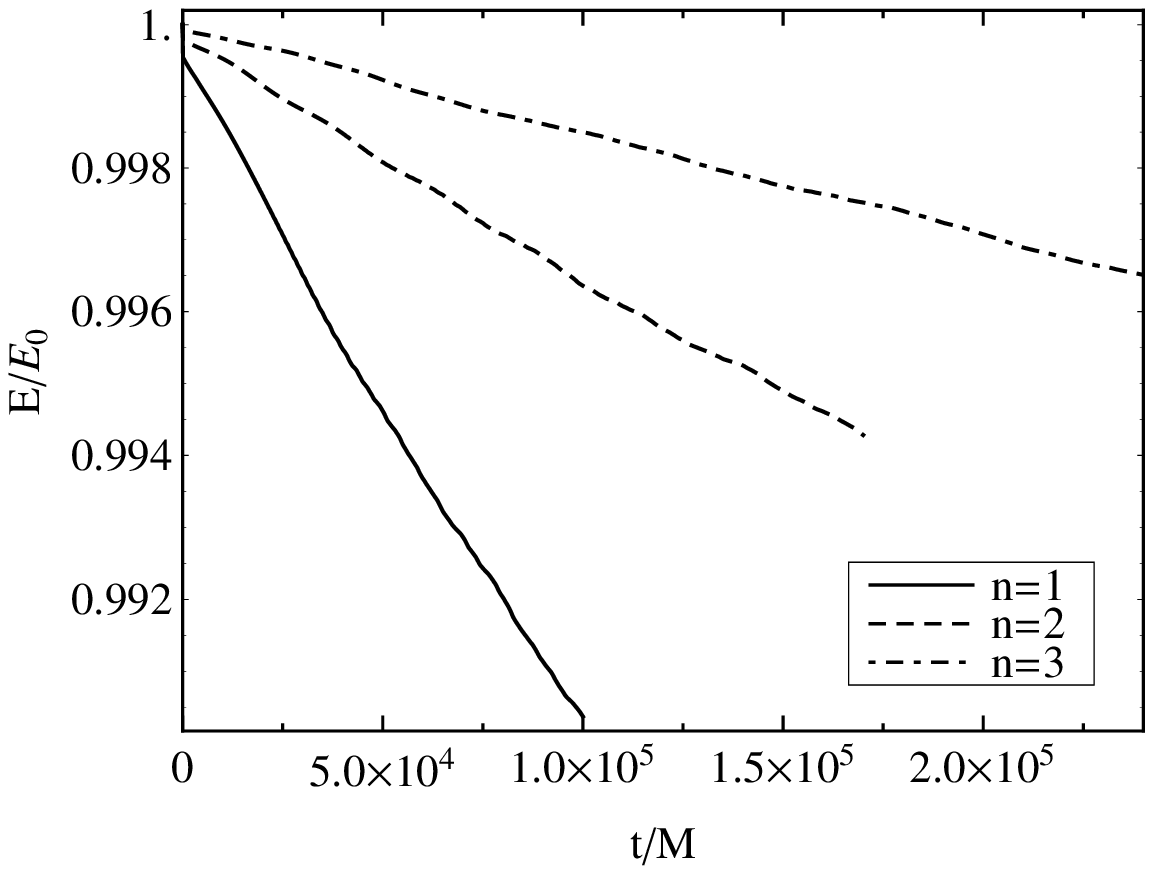}
\caption{Energy of the Dirac field for the evolution of pseudo-resonant initial data with $Mm=0.15$, $|\kappa|=1$, and different resonance modes ($n=1,2,3$). The upper figure shows the case with positive $\kappa$ and the lower one shows the case with negative $\kappa$. We can find that the evolution lasts longer for a bigger $n$, and for the same resonant mode, the time of evolution lasts much longer in the case with negative $\kappa$ .}
\label{Evolution1}
\end{center}
\end{figure}

\begin{figure}[htbp]
\begin{center}
\includegraphics[width=3in]{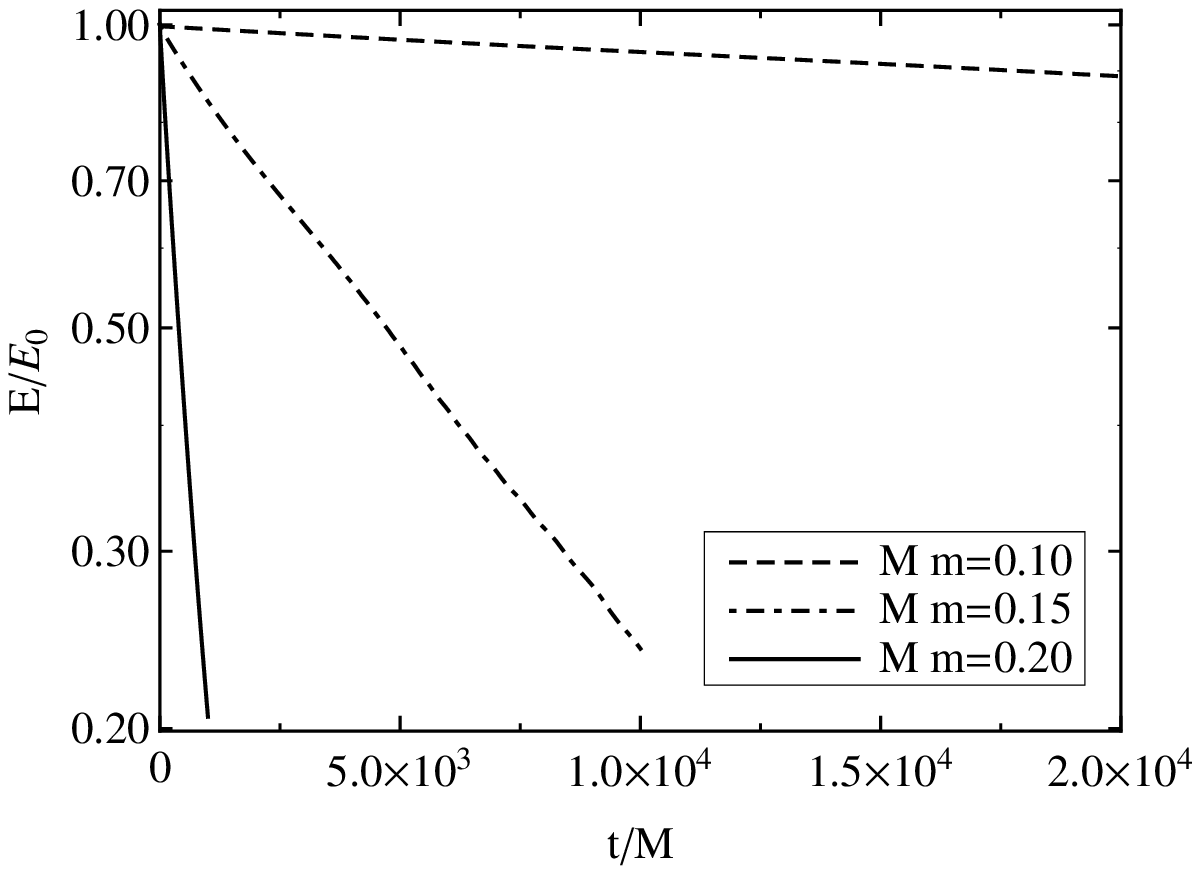}
\includegraphics[width=3in]{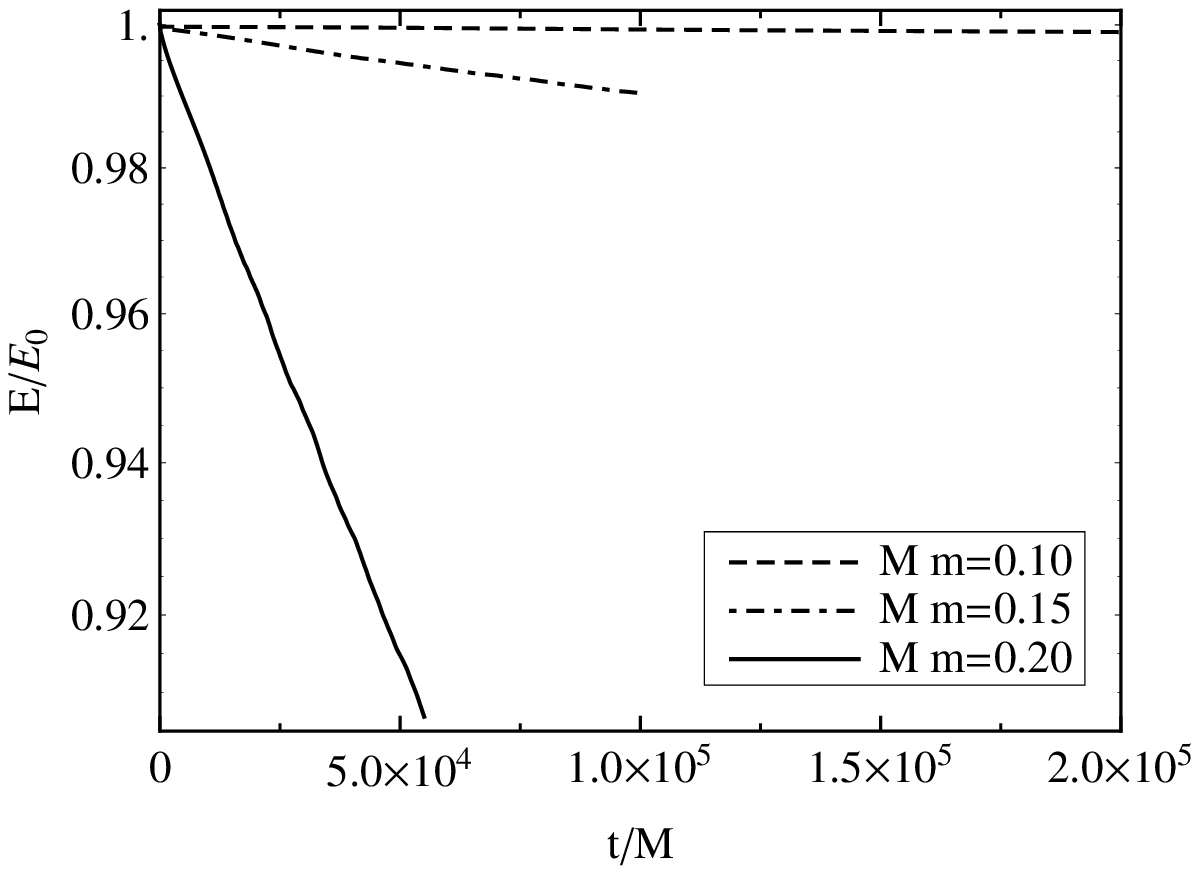}
\caption{Energy of the Dirac field for the evolution of pseudo-resonant initial data with $|\kappa|=1$, $n=1$, and different values of $Mm$. The upper figure shows the case with positive $\kappa$ and the lower one shows the case with negative $\kappa$. We can find that the evolution lasts longer for a smaller mass $Mm$. And for the same mass $Mm$, the evolution lasts much longer in the case with negative $\kappa$.}
\label{Evolution2}
\end{center}
\end{figure}

\begin{figure}[htbp]
\begin{center}
\includegraphics[width=3in]{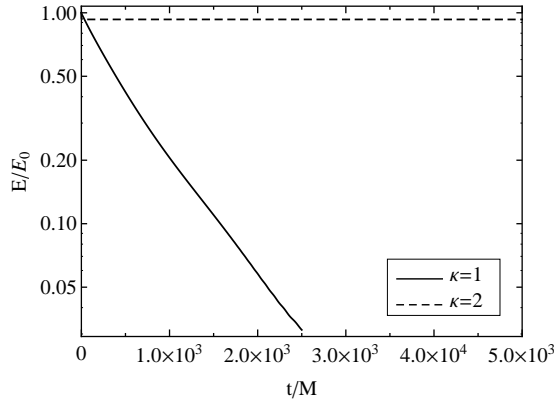}
\end{center}
\caption{Energy of the Dirac field for the evolution of pseudo-resonant initial data with $Mm=0.20$, $n=1$, and different values of $|\kappa|$. It is obvious that the evolution with a bigger value of $|\kappa|$ lasts much longer.}
\label{Evolution3}
\end{figure}

First we focus on the evolution of the pseudo-resonant initial data whose frequencies are given in Tab.~\ref{Resonantpoint}. Some of our results are shown in Fig.~\ref{Evolution1} and Fig.~\ref{Evolution2}. We can find that for the same $\kappa$ and $Mm$, the evolution lasts longer for the case with a larger resonant frequency. At the same time, for the same $\kappa$ and the same ordinal number $n$ of the resonant frequencies, the evolution with a smaller mass $Mm$ lasts longer. Similar conclusions can be seen in the scalar case, but when we consider the effect of the parameter $\kappa$, we get some interesting results. For the same value of $|\kappa|$ and other parameters, the evolution of the case with negative $\kappa$ lasts much longer than the one of the case with positive $\kappa$. This can be explained by the analysis of the effective potential as mentioned before. For the same $Mm$, $n$, and $|\kappa|$, the peak of potential is higher in the case with negative $\kappa$. Thus it is more difficult for particles in the potential well to tunnel through the potential barrier and fall into the black hole. As we know, the sign of $\kappa$ is related to the spin-orbit interaction. For positive $\kappa$, $\kappa=j+\frac{1}{2}$, and for negative $\kappa$, $\kappa=-\left(j+\frac{1}{2}\right)$. Obviously, the sign of $\kappa$ could represent the different ways of spin-orbit interaction for a same $j$. Thus the above discuss implies that the way of spin-orbit interaction of Dirac particles affects their evolution around a black hole, even if the black hole do not contain any charge. The value of $|\kappa|$ also affects the duration of evolution. Fig.~\ref{Evolution3} shows the evolutions of the first resonant mode with the same mass $Mm=0.20$ but different $|\kappa|$. We can find that for bigger $|\kappa|$, the evolution lasts much longer, which means that the Dirac particles with larger total angular momentum will be more likely to stay around a Schwarzschild black hole.

Our results show that the energy decays exponentially as excepted: $E=E_{0}\text{exp}(-st/M)$. Performing a linear fit of $\ln(E/E_{0})$, we can obtain the parameter $s$. The results are shown in Tab. \ref{slope_s}. Here we should note that if we change both the signs of $\omega$ and $\kappa$ for initial data, Eqs. (\ref{AG}) and (\ref{AF}) will give different solutions, but we find that the parameter $s$ still remains unchanged. So we only consider two cases: $M\omega \kappa>0$ and $M\omega \kappa<0$.

\begin{table}[htbp]
\caption{The slope $s$ of a linear fit of $\ln(E(t)/E_{0})$ with different $Mm$ and $n$. Here we set $|\kappa|=1$ and only consider two cases: $M\omega\kappa>0$ and $M\omega\kappa<0$.}
\label{slope_s}
\begin{tabular}{cccccc}
\hline\hline
        $~~~~M m~~~~$  &~~~~~~~~& $~~~~~0.10~~~~~$ & $~~~~~0.15~~~~~$ & $~~~~~0.20~~~~$ \\
\hline
                &$n=1$ & $5.6\times10^{-6}$ & $1.4\times10^{-4}$ & $1.6\times10^{-3}$ \\
$M\omega\kappa>0$    &$n=2$ & $7.8\times10^{-7}$ & $1.9\times10^{-5}$ & $2.1\times10^{-4}$ \\
               &$n=3$ & $2.0\times10^{-7}$ & $5.9\times10^{-6}$ & $6.0\times10^{-5}$ \\
\hline
               &$n=1$ & $3.4\times10^{-9}$ & $9.3\times10^{-8}$ & $1.7\times10^{-6}$ \\
$M\omega\kappa<0$    &$n=2$ & $1.1\times10^{-9}$ & $3.3\times10^{-8}$ & $6.4\times10^{-7}$ \\
               &$n=3$ & $4.9\times10^{-10}$ & $1.4\times10^{-8}$ & $2.8\times10^{-7}$ \\
\hline\hline
\end{tabular}
\end{table}

\begin{figure}[htbp]
\begin{center}
\includegraphics[width=3in]{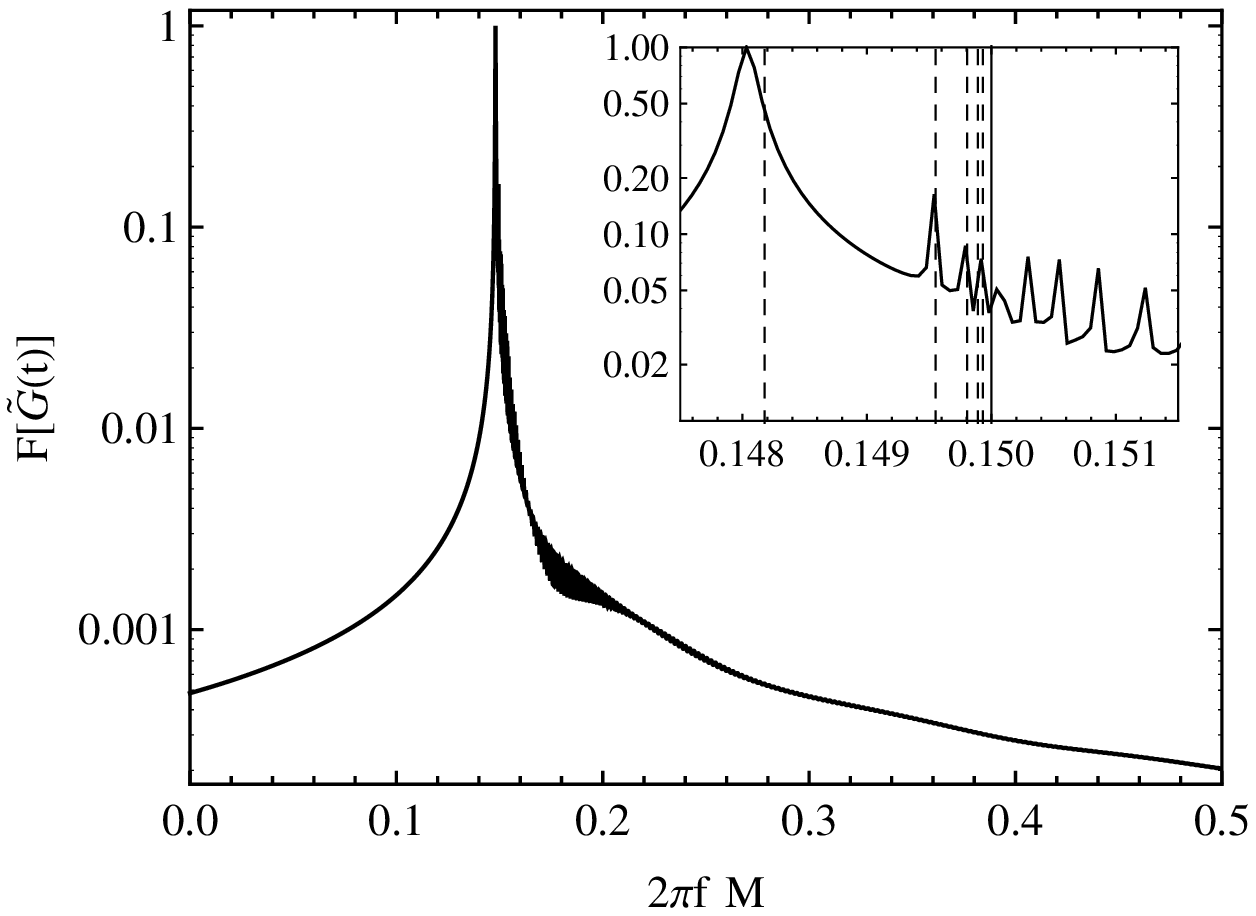}
\includegraphics[width=3in]{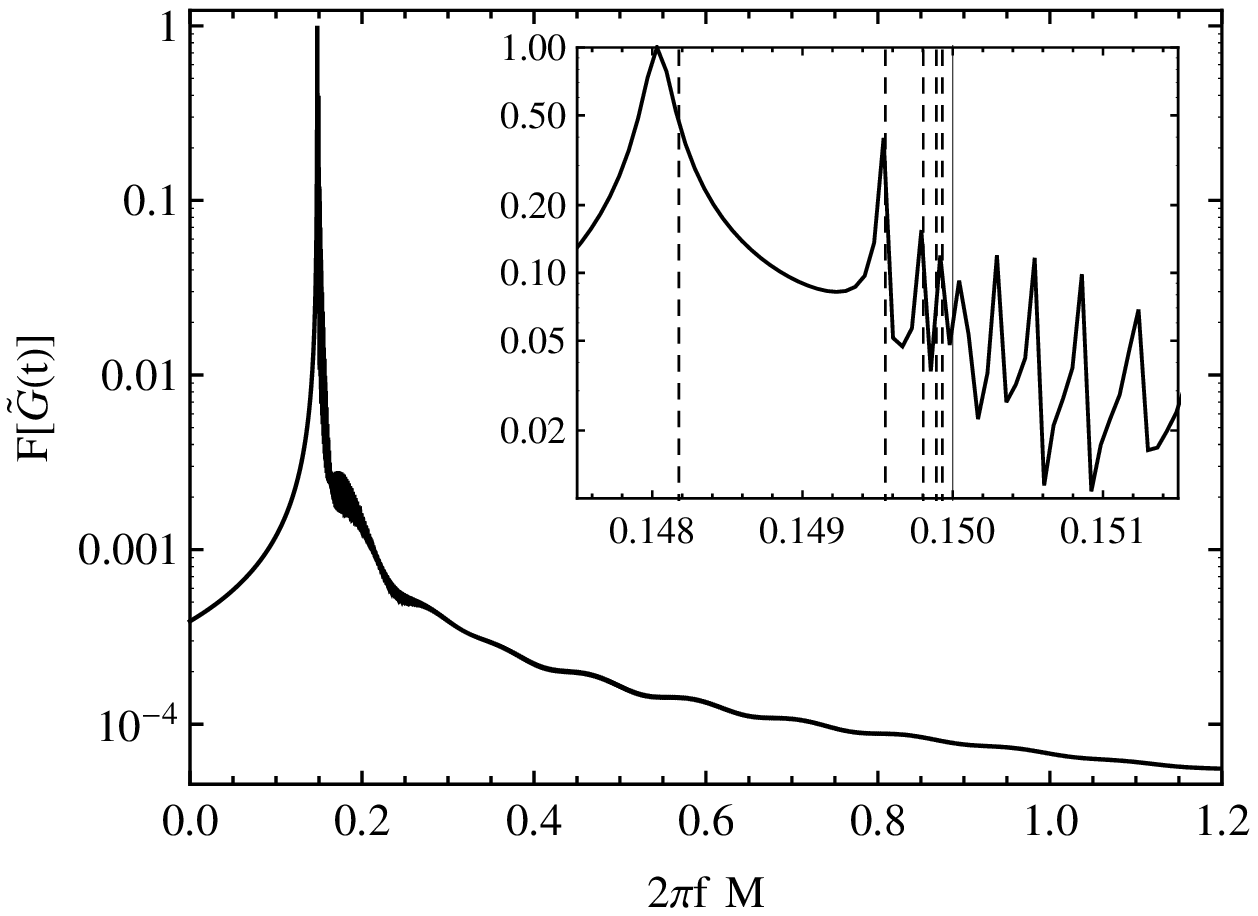}
\caption{Discrete Fourier transform in time for the evolution of different initial data: pseudo-resonant initial data (upper plot) and non-resonant initial data (lower plot). Here we have chosen $\kappa=1$ and $Mm=0.15$. For the upper figure, $M\omega$ is set to 0.1481772 corresponding to the first pseudo-resonant mode. For the lower figure, $M\omega$ is set to 0.1486650 as a comparison. The overlay figures show more details of the region near the first peak. The dashed vertical lines in the overlay figures denote the first five resonant frequencies.}
\label{Fourier}
\end{center}
\end{figure}

\begin{figure}[htbp]
\begin{center}
\includegraphics[width=3in]{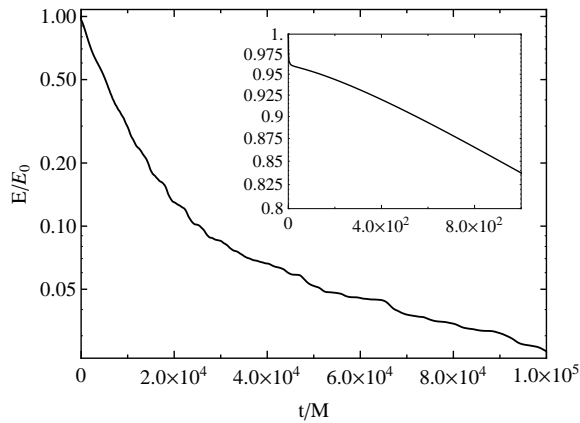}
\end{center}
\caption{Energy of the Dirac filed for the evolution of non-resonant initial data with $\kappa=1$, $Mm=0.15$, and $M\omega=0.1486650$. The overlay figure shows more details of the early evolution.}
\label{Evolutionofnonresonant}
\end{figure}

To understand more details of the process of evolution, we perform a spectral analysis as in Ref. \cite{PhysRevD.84.083008}. We calculate the discrete Fourier transform in time of the Dirac field at a fixed point $r_j$. The magnitude of the discrete Fourier transform is
\begin{equation}
F[\tilde{G}(t)](f):=\left|A\sum_p\tilde{G}(t_p, r_j) \text{exp}(-2\pi\text{i}f t_p)\right|,\label{FF}
\end{equation}
where $A$ is a normalization constant and $t_p$ are the discrete time values. Note that we can also perform the discrete Fourier transform on $\tilde{F}$, which will give the same results. The results for different initial data with $\kappa=1$ and $Mm=0.15$ are shown in Fig.~\ref{Fourier}. We can find that there exist clear peaks at different resonant frequencies. Similar conclusions can be seen in the scalar case
Ref.~\cite{PhysRevD.84.083008}. The upper figure in Fig.~\ref{Fourier} shows the case of pseudo-resonant initial data with  $M\omega=0.1481772$ corresponding to the first pseudo-resonant mode. We can see a clear peak near the first resonant frequency. The lower figure in Fig.~\ref{Fourier} shows the case of non-resonant initial data with $M\omega=0.1486650$, which is between the first and second resonant frequencies. We can see two comparable peaks near the first and second resonant frequencies. Furthermore,  in both cases there exist small peaks near other resonant frequencies. These analyses imply that the non-resonant states will evolve as a combination of resonant states after a very shot time. This can also be seen from the energy of the Dirac filed for the evolution of non-resonant initial data (see Fig.~\ref{Evolutionofnonresonant}). It is obvious that there are three stages through the evolution: a) the energy falls off quickly in a very short time; b) then it decays exponentially for a long time; c) finally it turns into a state of power-law damping. The last stage is the so-called power-law tail behavior at very late time \cite{PhysRevD.87.043513,PhysRevD.72.027501}. This process is very similar to the one in the Dirac quasinormal mode \cite{PhysRevD.72.027501}. For the scalar case, the authors of Ref.~\cite{PhysRevLett.109.081102} investigated and showed the relations between the scalar stationary resonances, quasinormal modes, and the dynamical resonance states. One can see that the scalar dynamical resonance states naturally yield a similar role to the quasinormal modes. Similar conclusions may apply to the Dirac field, which needs to be discussed in detail in future works.

At last, we consider the half-life time of the Dirac dynamical resonance states. It is defined as $t_{1/2}=\ln(2)M/s$. For a Schwarzschild black hole with mass $M=10^{8}M_{\bigodot}$, the Dirac dynamical resonance state with $s=4.9\times10^{-10}$ (which corresponds to $|\kappa|=1$, $Mm=0.1$, $n=3$, and $M\omega\kappa<0$), $t_{1/2}$ will reach 44000 years. Note that the half-life time will increase with decreasing $Mm$. In Ref.~\cite{PhysRevD.84.083008}, the authors argued that if we consider ultra-light scalar field, the evolution could last for cosmological time-scales. In the Dirac case, the parameter $s$ also decreases with decreasing $Mm$. Thus for ultra-light Dirac field, particles can stay in the potential well for a long time, even for cosmological time-scales. In addition, we find the parameters $\kappa$ and $\omega$ also affect the lifetime of evolution, i.e., the parameter $s$ decreases with increasing $|\kappa|$; and for the same values of $|\kappa|$ and $|\omega|$, the parameter $s$ in the case $\kappa\omega<0$ is much smaller.

\section{Conclusion and discussion}
\label{section5}

In this paper, we have investigated the Dirac dynamical resonance states and their evolutions around a Schwarzschild black hole by numerical method. First, we considered the Dirac equation in the Schwarzschild black hole spacetime and obtained two Schr\"{o}dinger-like equations, which we used to investigate the behavior of the Dirac field around the black hole. The corresponding two effective potentials are supersymmetric partners, hence the spectra are of the same. Each effective potential $V$ depends on three parameters: the mass of the Dirac field $m$, the parameter $\kappa$ associated with spin-orbit interaction, and the oscillating frequency $\omega$. Here $\kappa$ goes over all positive and negative integers. We found that there exists ``potential well" in the effective potential for some special parameters, and resonant states appear when the frequency $\omega$ lies in the resonance band, i.e., $V^{min}<\omega^2<\text{min}\{V^{max},m^2\}$. However, the specific form of the effect potential $V$ is so complex that it is difficult to solve the equation $\partial_{r}V(r)=0$ analytically to obtain the extremums of the potential. At the same time, $V$ depends on $\omega$.
Thus we looked for the resonance band by numerical method and showed our results in Fig.~\ref{band}. We found that the cutoff of $Mm$ increases with $|\kappa|$ and is larger when $\kappa<0$. And the resonance region of $\omega$ becomes narrower and narrower when $Mm$ approaches 0.

Then we solved the Schr\"{o}dinger-like equation (\ref{EF}) numerically to obtain the resonant frequencies for different parameters. The results were shown in Fig.~\ref{Resonant} and Tab.~\ref{Resonantpoint}. There seem always exist infinite numbers of resonant states because when $M\omega\rightarrow Mm$, the resonant frequencies become closer and closer to each other. The resonant spectra are almost the same for the same value of $|\kappa|$, except for the first resonant frequency of the case with positive $\kappa$.

Using these stationary resonant solutions, we constructed the configurations of the Dirac dynamical resonance states as initial data, and investigated their evolutions and energy decay. The results were shown in Figs.~\ref{Evolution1}, \ref{Evolution2}, and \ref{Evolution3}. As in the scalar case, the energy of the Dirac dynamical resonance states shows an exponential decay. For the same $\kappa$ and $Mm$, when the resonant frequency $M\omega_{n}\rightarrow Mm$, the evolution lasts longer. On the other hand, the lifetime of evolution increases with decreasing $Mm$. The effect of $\kappa$ is very interesting. The lifetime of evolution increases with $|\kappa|$ and if we consider the same $Mm$, $n$, and $|\kappa|$, the evolution lasts much longer when $\kappa<0$. The reason is obvious if we consider the effect of $\kappa$ on the effective potential $V_{1}$ (\ref{massiveV}). As mentioned before, the peak of potential increases with the value of $|\kappa|$ and for the same $Mm$, $n$, and $|\kappa|$, the peak of potential will be higher in the case with negative $\kappa$. As the potential barrier becomes higher, it will be more difficult for particles in the potential well to tunnel through the potential barrier. So they can stay longer around a black hole. The value of $|\kappa|$ is related to the total angular momentum number $j$ and the sign of $\kappa$ could represent the way of spin-orbit interaction. This implies that not only the total angular momentum, but also the way of spin-orbit interaction of Dirac particle affects the lifetime of its evolution around a black hole, even if the black hole is an uncharged black hole. The lifetime of evolution can be described by a parameter $s$ which can be obtained by a linear fit of $\ln(E/E_{0})$ as the function of $t/M$. The results were shown in Tab.~\ref{slope_s}. It is worth noting that, in Ref.~\cite{Lasenby:2002mc}, A. Lasenby, C. Doran, J. Pritchard, A. Caceres and S. Dolan have discussed the dacay of a very similar kind of Dirac modes. After some private communication with S. Dolan, a good agreement was found on the numerical results. The possible relation will be discussed further in other place.

For the non-resonant initial data, the behavior of evolution could be treated as a combination of evolutions of different pseudo-resonant initial data after a very short time. There are three stages in the evolution. At the first stage, the energy falls off very quickly in a very short time. At the second stage, the energy decays exponentially for a along time. Finally at the last stage, the energy goes through a power-law damping at very late time, which is called the power-law tail behavior \cite{PhysRevD.72.027501}. Then we considered the half-life time of the Dirac dynamical resonance states and got the similar conclusion as in the scalar case: for ultra-light Dirac field, particles can stay around a Schwarzschild black hole for a very long time, even for cosmological time-scales. In addition, considering the effect of $\kappa$, the result will be much richer. In fact, according to Ref.~\cite{PhysRevLett.109.081102}, the scalar dynamical resonance states can be described by a state with complex frequency: the real part of the frequency is the same with the frequency of the stationary resonance and the imaginary part is related to the quasinormal mode and represents the rate of decay. We suppose it would also apply to the Dirac case and will discuss this in detail in future works.

There are still some questions. Firstly, in this paper we just considered a test Dirac field in the Schwarzschild spacetime, if the back reaction is also considered, we should get more interesting results. Secondly, the more realistic black holes are Kerr black holes, so we should investigate the Dirac dynamical resonance states around Kerr black holes in our future works. Lastly, we can also consider the coupling of the Dirac field and scalar field in a black hole spacetime.

\section*{Acknowledgement}
The authors thank Prof. J. Barranco for his kind help and answers on the initial conditions of the numerical evolution and S. Dolan for comparing the results in their paper ~\cite{Lasenby:2002mc} with ours.  XLD would like to thank Prof.  Xin-He Meng for beneficial discussions during the seminar of Planck satellite and AMS held in KITPC.
This work was supported in part by the National Natural
Science Foundation of China (Grant No. 11075065 and Grant No. 11375075), the Huo Ying-Dong Education Foundation of
Chinese Ministry of Education (Grant No. 121106), and the Fundamental Research Funds for the
Central Universities (Grant No. lzujbky-2013-18). K. Yang was supported
by the Scholarship Award for Excellent Doctoral Student granted by Ministry of
Education.

\end{document}